\documentclass[aps,pra,reprint,longbibliography]{revtex4-1}
\usepackage{graphicx}
\usepackage{amsmath}
\usepackage{amssymb}
\usepackage{amsfonts}
\usepackage{dcolumn}

\begin{document}
\title{Overcoming experimental limitations in a non-linear two-qubit gate through postselection}
\author{Juli\'{a}n Mart\'{i}nez-Rinc\'{o}n}
\email{jmarti41@ur.rochester.edu}
\affiliation{Department of Physics and Astronomy \& Center for Coherence and Quantum Optics, University of Rochester, Rochester, New York 14627, USA}
\date{\today}

\begin{abstract}
We introduce a modular-value two-qubit gate and explore its advantages in experimentally-limited situations. The gate is defined such that the final state of a qubit is fully controlled by a pre- and post-selection procedure in an ancillary qubit given an (imperfect or technically limited) intermediate conditional qubit-qubit interaction. As an example of the gate and its benefit, we make the connection to a postselected cross phase modulation scenario when a undesired absorption is present. Probabilistic amplification of a small cross phase while mitigating relative absorption is possible, and a complementary behaviour between phase and amplitude emerges. 
\end{abstract}

\maketitle 
\section{Introduction}
Weak-value amplification (WVA) is a metrological precision technique for estimation of small parameters that allows one to amplify a signal of interest above a technical-noise floor. Such a response is possible with a unorthodox post-selection procedure in a system after weakly coupling it to a continuous degree of freedom or meter. The technique's advantages rely in the anomalous shift induced in the meter, which is proportional to the weak value of the system's observable. The weak value is a unbounded complex quantity which emerges as the result of inducing a weak system-meter interaction and collecting the events surviving the post-selection. For details about how the technique is implemented and its technical-noise mitigation advantages see review papers~\cite{Understanding,ReviewNori,ReviewShikano,ReviewSvensson} and  references~\cite{TorresWVA,Brunner,Starling,singlephotonnonlinear,PRX,NoiseExp}.   



A different approach than WVA to use a weak value has been done when exchanging the dimensionality of the system and the meter, i.e. by choosing the system as a continuous degree of freedom and the meter as a qubit. Such an idea has allowed researchers to make the weak value directly proportional to the wave function of the system, which can then be directly measured by tracking the post-selection surviving meter qubits~\cite{MeasuringWaveFunction,DirectMeasurement,MehulDirect}. Such a protocol offers an easier alternative to quantum tomography by directly characterizing a state. However it has been recently claimed that the weak interaction condition is not required nor desired for performing such a task~\cite{DirectStrongChina,DirectStrongItaly}, leaving postselection on an ancillary system as the principal requirement. Qubit state direct tomography using weak values has also taken some recent interest~\cite{SalvailPolarization,2016arXiv160604659Q}. 

We are interested here in the case where both, system and meter, are qubits. A two-qubit system was used to reproduce some of the original ideas of Aharonov \textit{et.al.}~\cite{AAV} for the first time few years after the original proposal~\cite{Suter}, later for separate parties using two-photon entanglement~\cite{PhotonPolarization}, also to prove violation of Leggett-Garg-type inequalities in optical systems~\cite{Leggett-GargPryde,Leggett-GargJustin} and superconductive circuits~\cite{SuperconductiveDiCardo,Martinis2016}, and more recently in the matter-wave interferometry experiments~\cite{Denkmayr2014,WVmatter-wave,CheshireCatBrazil}. In addition, it has been proposed the direct measurement and study of the quantum weak value using two qubits~\cite{BrunGate,Molmer,Cormann}, and also the possibility of using two-qubit gates based on weak measurements for computation~\cite{Lund,Laflamme}. On the other side, postselection alone has been used to overcome non-Markovian dephasing in single-photon sources~\cite{Nazir}. We will address in this paper the ability to \textit{fully} control the final state of a qubit (system) by locally postselecting an ancillary qubit after entangling both of them. We will show that such a protocol allows us to induce a WVA-like response in the two-qubit scenario in order to overcome imperfections or technical limitations in experiments. Our protocol could be used within measurement-based quantum computation schemes~\cite{Measurement-basedQC}, where highly entangled states are required to exploit non-classicality for information processing. Our proposed gate is designed to use postselection to overcome technical or experimental difficulties to develop non-linear two-qubit gates, and it is independent of the degree of qubit-qubit correlations in the gate. For the same reason, schemes such as KLM~\cite{KLM} for linear computation are excluded from our protocol.

The technique of WVA requires a weak interaction between the system and the meter to satisfy the requirement of small disturbance in the system's state. When considering two qubits, however, the probability for the interaction to occur can be controlled with the state preparation of the qubits, which determines the response after a given two-qubit gate. This difference was noted by Kedem and Vaidman~\cite{ModularValues} whose introduced the \textit{Modular Value} to describe the properties of a post-selected two-qubit system under an interaction of any given strength. Also, unlike the WVA protocol where the imaginary and the real part of the weak value independently determine the shifts in two conjugate continuous variables~\cite{jozsa}, separating the modular value in its magnitude and phase allows for a better understanding of the qubit's dynamics~\cite{Cormann}. We introduce in this work a postselection-based two-qubit gate where a qubit's final state is defined by the modular value for any given local operation on an ancillary qubit (Section ~\ref{SectionGate}). As an example, we show how such a modular value-controlled gate can be used in a post-selected cross phase modulation scenario to mitigate a undesired experimental absorption (non-unitary evolution) and to amplify the effective cross phase. In addition, an interesting behaviour emerges such that either a relative phase can be controlled by a relative absorption plus post-selection, or a relative absorption can be controlled by a relative phase plus post-selection, or also that the absorption can be \textquotedblleft translated" to a different eigenstate in a coherent superposition (Section ~\ref{SectionXPM}). 
 

\section{Modular value-controlled gate}\label{SectionGate}

\begin{figure}[h]
\centering
\includegraphics[width=0.45\textwidth]{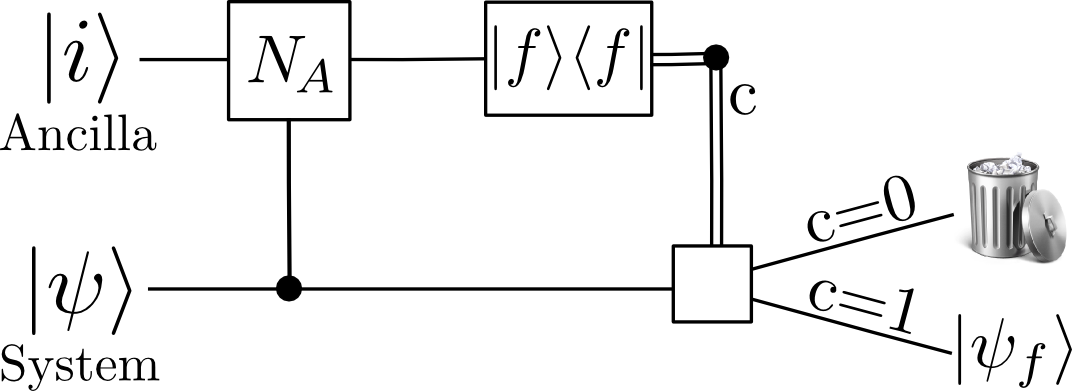}
\caption{Circuit representation of a two-qubit modular value-controlled gate. Given an intermediate qubit-qubit interaction $\hat{N}_{S,A}$, preselection and postselection of the ancilla qubit offers full control on the postselection-succeeded final state of the system qubit, $|\psi_f\rangle$. The classical bit $c=\{1,0\}$ determines if the postselection is successful or not respectively, such that if $c=0$ the system qubit is discarded. 
}
\label{fig:protocol}
\end{figure}

Following the general definition given in Ref.~\cite{Cormann} we consider the two-qubit gate pictured in Fig.~\ref{fig:protocol}. An ancillary qubit and a system qubit are prepared in the known states $|i\rangle$ and $|\psi\rangle$ respectively. A given single-qubit operator $\hat{N}_A$ acts on the ancillary qubit only if the system qubit is in the eigenstate $|1\rangle$.
The global operator acting on the qubits takes the form,
\begin{equation}\label{NTC}
\hat{N}_{S,A}=|0\rangle\langle0|_S\otimes \hat{I}_A+|1\rangle\langle1|_S\otimes \hat{N}_A,
\end{equation}   
where the subindices $S$ and $A$ label the system and the ancillary qubit subspaces respectively, $\{|0\rangle,|1\rangle\}$ is the system's basis, and $\hat{I}$ is the identity operator. Such an interaction can also be expressed in the measurement evolution of the form $\hat{N}_{S,A}=\exp{(-ig\hat{O}_A\otimes |1\rangle\langle 1|_S)}$, where $\hat{N}_A=\exp{(-ig\hat{O}_A)}$, $\hat{O}_A$ is an operator on the ancillary qubit, and $g$ is the strength of the ancilla-system interaction. 
We call \textquotedblleft system" the qubit we are interested to control via pre- and post-selection of the ancillary qubit. The operation $\hat{N}_A$ is applied onto the ancillary qubit conditioned to the state of the system qubit, as in Eq.~(\ref{NTC}). 

After the qubit-qubit interaction, the ancillary qubit is postselected on the state $|f\rangle$, and the final system qubit state is recorded only if the postselection is successful. The final (unnormalized) state for the system qubit is given by, 
\begin{eqnarray}\label{final}
|\tilde{\psi}_f\rangle&=&\langle f|\hat{N}_{S,A}\left(|\psi\rangle\otimes|i\rangle\right)\nonumber\\
&=&\langle f|i\rangle\left[\cos(\theta/2)|0\rangle+N_m\,e^{i\xi}\sin(\theta/2)|1\rangle\right],
\end{eqnarray}
where we have assumed an initial state for the system qubit given by 
\begin{equation}
|\psi\rangle=\cos(\theta/2)|0\rangle+e^{i\xi}\sin(\theta/2)|1\rangle.
\end{equation}
Here,
\begin{equation}
N_m=\frac{\langle f|\hat{N}_A|i\rangle}{\langle f|i\rangle}
\end{equation} 
is the modular value of $\hat{N}_A$~\cite{ModularValues}. Even though $N_m$ has the same form than the weak value of $\hat{N}_A$ for a given pre- and post-selection choice, the two values have fundamental differences. The weak value of a qubit observable is defined when postselected after a weak interaction with other degree of freedom (of any dimensionality). The modular value is defined for a qubit-qubit coupling of any given strength, so the result in Eq.~(\ref{final}) is valid for any single-qubit operator $\hat{N}_A$ given the global interaction of Eq.~(\ref{NTC}). 

The protocol effectively gives a relative factor $N_m$ between the basis states of the initially prepared state $|\psi\rangle$ (see eq.~\ref{final}). This means that the final state of the system qubit can be controlled by pre- and post-selection on the ancillary qubit. Such a control is accompanied by a success probability given by
\begin{eqnarray}
p_{N_{S,A}}&=&\langle\tilde{\psi}_f|\tilde{\psi}_f\rangle\nonumber\\
&=&|\langle f|i\rangle|^2\,\left[\cos^2(\theta/2)+|N_m|^2\sin^2(\theta/2)\right], 
\end{eqnarray}
which depends on the overlapping of the pre- and post-selection states of the ancillary qubit $|\langle f|i\rangle|$, the preparation of the system qubit $|\psi\rangle$, and the modular value of $\hat{N}_A$. Note that based on the choice of the local operator $\hat{N}_A$ one can in principle couple the qubit pair in any possible way with any given strength. However if the initial state of the system is $|\psi\rangle=|0\rangle$, for example, both ancilla and system remain unchanged after passing through the gate, independently of the choice for $\hat{N}_A$. For this reason, it is usually common to refer to $\theta$ as the \textquotedblleft interaction strength" or the \textquotedblleft probability of interaction'' in this situations~\cite{Lund,Cormann}. We propose our protocol for scenarios where the realization of $\hat{N}_A$ is imperfect or does not satisfy a desired goal due to experimental limitations when operating the qubit-qubit interaction of Eq.~(\ref{NTC}).

The normalized final state for the system qubit takes the form,
\begin{eqnarray}\label{finalnormalized}
|\psi_f\rangle&=&|\tilde{\psi_f}\rangle/\sqrt{p_{N_{S,A}}}\nonumber\\
&=&\cos\left(\frac{\theta-\theta_m}{2}\right)|0\rangle+e^{i(\xi+\Omega_m)}\sin\left(\frac{\theta-\theta_m}{2}\right)|1\rangle,
\end{eqnarray} 
where $\Omega_m$ is the argument of the modular vale, $N_m=|N_m|e^{i\Omega_m}$, and 
\begin{equation}\label{thetam}
\tan\left(\frac{\theta_m}{2}\right)=\frac{(1-|N_m|)\tan(\theta/2)}{1+|N_m|\tan^2(\theta/2)}.
\end{equation}
The phase $\Omega_m$ can in principle take any value within the interval $[0,2\pi)$, and the induced azimuthal rotation behaves as $\theta_m\approx\theta$ for $|N_m|\ll1$, $|\theta_m|\ll1$ for $|N_m|\approx1$, and $\theta_m\approx\theta-\pi$ for $|N_m|\gg1$. 

The result in Eq.~(\ref{finalnormalized}) is exact and general for any operator $\hat{N}_A$, the given two-qubit controlled gate operation of Eq.~(\ref{NTC}), the preselection $|i\rangle$, and the postselection $\langle f|$ in the ancillary qubit. As a result, the two-qubit full operation ($\hat{N}_{S,A}$ plus postselection) effectively introduces two rotations on the Bloch sphere representation of the postselection-surviving system qubit's state: one about the z-axis, given by $\Omega_m$, and one along the azimuthal direction, given by $-\theta_m$. The magnitude of these rotations on the system qubit are controlled by the modular value $N_m$ of the ancillary qubit.

We note that such an operation of Eq.~(\ref{finalnormalized}) on the initial state $|\psi\rangle$ can alternatively be deterministically obtained by consecutively applying two single-qubit rotations and without the need of the ancillary qubit. Such a procedure consists in rotating the system qubit about the y-axis by an angle $-\theta_m$, and then about the z-axis by an angle $\Omega_m$, $|\psi_f\rangle=R_z(\Omega_m)R_y(-\theta_m)|\psi\rangle$, with unity probability of success. However, such a procedure is local on the system qubit and does not follow the cross effect of a two-qubit gate. We take a different fundamental approach to our proposed gate by noticing that the control obtained due to post-selection can be used to overcome experimental limitations in non-linear two-qubit gates. We show an example of this advantage in the next section by first noticing that a similar result to Eq.~(\ref{finalnormalized}) can also be obtained from applying one imperfect (nonunitary) single-qubit phase shift gate, instead of two unitary rotations, on the system qubit. By introducing the ancillary qubit, we then show that optimization of cross phase modulation is possible and that complementarity between phase and amplitude emerges for such a case.

\section{Cross Phase Modulation conditioned to post-selection on the control photon}\label{SectionXPM}

\begin{figure}[h]
\centering
\includegraphics[width=0.47\textwidth]{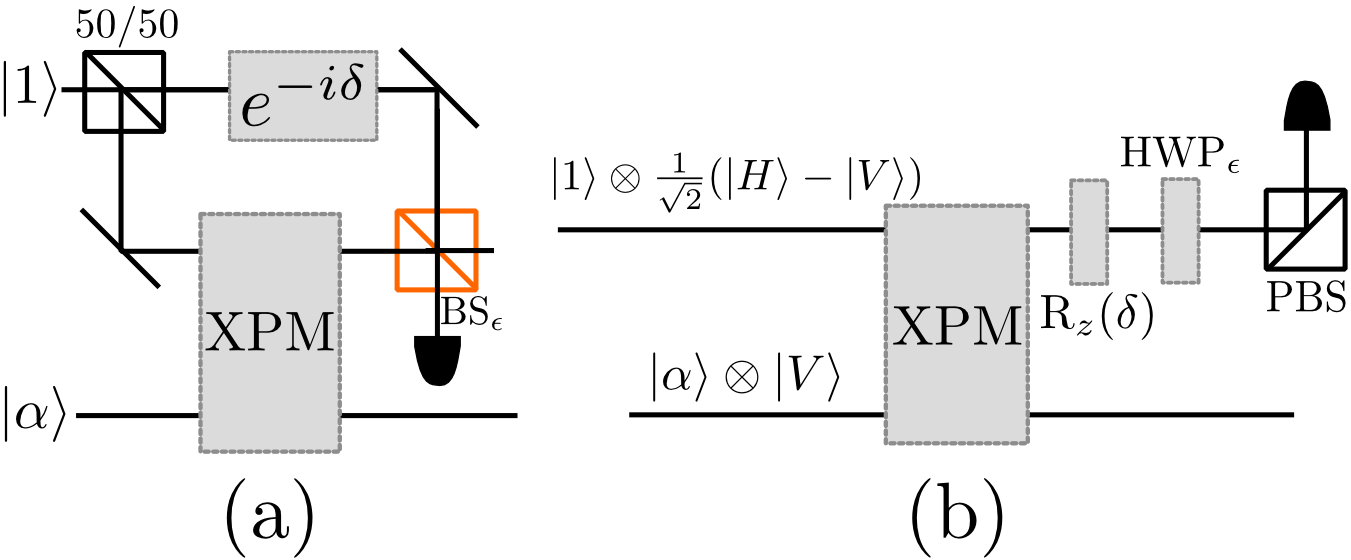}
\caption{Cross phase modulation (XPM) conditioned to postseletion on the control photon. The cross phase is imparted onto the weak coherent state $|\alpha\rangle$, and its final state is recorded only when the control photon is detected. Postselection can be performed using the which-path degree of freedom in an interferometer (a), or using polarization (b). In case (a), the single photon enters the interferometer through a 50/50 beam-splitter and postselection is controlled by introducing a relative phase $\delta$ between both arms and by using a recombining unbalanced beam-splitter, BS$_\epsilon$. In case (b), the control photon is prepared in the anti-diagonal linearly polarization state, and the XPM is induced only in the vertical component of polarization. Postselection is controlled by a combination of a phase shifter $R_z(\delta)$, a half-wave plate (HWP$_\epsilon$), and a polarized beam-splitter (PBS).}
\label{fig:XPM}
\end{figure}

We show in this section that Cross Phase Modulation (XPM) conditioned to postselection is a direct example of a modular value-controlled gate. Consider a single photon that imparts a conditional phase shift onto a coherent state $|\psi\rangle=|\alpha\rangle$, as shown in Fig.~\ref{fig:XPM}. We assume a highly attenuated coherent laser beam, $|\alpha|\ll1$, such that $|\psi\rangle\approx|0\rangle+\alpha|1\rangle$, where $\theta\approx2|\alpha|$, and the system qubit is defined by restricting the interaction to the vacuum state $|0\rangle$ and the one-photon Fock state $|1\rangle$.

The preselected state for the control photon is given by 
\begin{equation}
|i\rangle=\frac{1}{\sqrt{2}}\left(|-\rangle-|+\rangle\right),
\end{equation}
where $\{|-\rangle,|+\rangle\}$ is the ancillary qubit's basis. Fig.~\ref{fig:XPM} shows examples on how the preselection and postselection can be done using either the which-path degree of freedom in an interferometer or the linear polarization of the control photon. The global operator~(\ref{NTC}) takes the form,
\begin{equation}\label{XPMeq}
\hat{N}_{XPM}=|0\rangle\langle0|\otimes\hat{I}+|1\rangle\langle1|\otimes \hat{R}_z^{NU}(\phi), 
\end{equation}
where $\hat{I}=|+\rangle\langle+|+|-\rangle\langle-|$, and only the one-photon component of the coherent state $|\alpha\rangle$ introduces the phase shift in the control photon (basis $\{|-\rangle,|+\rangle\}$).

We consider the cross phase shift to be limited by undesired experimental absorption, i.e.
\begin{equation}\label{nonunitary}
R_z^{NU}(\phi)=\begin{bmatrix}
1&0\\
0&e^{-a}\,e^{i\phi}\\
\end{bmatrix}\approx\begin{bmatrix}
1&0\\
0&(1-a)\,e^{i\phi}\\
\end{bmatrix},
\end{equation}
where $\phi$ is the conditional cross phase shift, $NU$ refers to the \textit{Non-unitarity} of the operation, and the parameter $a\ll1$ characterizes the small absorption of the non-linear gate. Our example here for the gate introduced in the previous section is for a weak interaction, were $2|g|=\sqrt{\phi^2+a^2}\ll1$ and $\hat{O}_A=\hat{\sigma}_z$. The small absorption allows us to approximate the noisy channel as the action of one single Kraus operator, giving by Eq.~(\ref{nonunitary}). The second Kraus operator of the channel is given by 
\begin{equation}
\begin{bmatrix}
0&0\\
0&\sqrt{1-e^{-2a}}\end{bmatrix}\approx\begin{bmatrix}
0&0\\
0&\sqrt{2a}\\
\end{bmatrix},
\end{equation}
which we ignore since $a\ll1$. Under this approximation, the phase shift plus small relative absorption is expressed as an imperfect but still unitary phase shift operation (see Appendix for further discussion).   

The rotation of Eq.~(\ref{nonunitary}) applied directly to the state $|\alpha\rangle$, instead of using our protocol of Eq.~(\ref{XPMeq}), would be the case where the ancillary degree of freedom for the control photon, which-path or polarization in Fig.~\ref{fig:XPM}, is not used (no post-selection). In such a scenario the single photon induces the cross phase shift $\phi$ in the state $|\alpha\rangle$, accompanied by an azimuthal rotation given by $\Delta\theta\approx -a\sin\theta$ (see Appendix~\ref{appendix}). Such undesired small rotation $\Delta\theta\approx -2a|\alpha|$ is usually seen as an error, and can in principle be corrected by rotating the qubit back in the azimuthal direction. Our approach is based on experimental restrictions, for example when the absorption $a$ is experimentally fixed for a given desired value of the cross phase $\phi$. By using the modular value-controlled gate (ancillary degree of freedom plus its postselection) in such imperfection-limited realizations, a qubit-qubit cross control can be used to overcome experimental limitations.  


We apply the global operation of Eq.~(\ref{XPMeq}) and by postselecting the final state of the ancillary qubit an \textit{effective} single-qubit gate is induced to the system qubit. Instead of the experimentally-fixed values $\phi$ and $\Delta\theta$ of the ancilla-less scenario, the system qubit is rotated by the controllable phases $\Omega_m(|i\rangle,|f\rangle;\phi,a)$ and $\theta_m(|i\rangle,|f\rangle;\phi,a)$ as in Eq.~(\ref{finalnormalized}). The postselection-surviving coherent state takes the form 
\begin{equation}
|\alpha_f\rangle\propto(|0\rangle+\alpha R_m|1\rangle),
\end{equation}
 where $R_m=\langle f|\hat{R}_z^{NU}|i\rangle/\langle f|i\rangle$ is the modular value of $\hat{R}_z^{NU}$.

\subsection*{Modular-Value Amplification}

A well-known result of the WVA protocol is to amplify a small interaction for ease detection. Such a result is possible under an almost-orthogonal pre- and post-selection choice. We show here a similar response using the modular value of $\hat{R}_z^{NU}$, where the cross phase $\phi$ is considered very small and can be arbitrarily amplified. As an example, this scenario could be implemented to improve the visibility of macroscopic state interferometry over large distances~\cite{Franson}, by amplifying a detector-limited very small cross phase. The postselection state is chosen as
\begin{equation}
|f\rangle=\frac{\cos\epsilon-\sin\epsilon}{\sqrt{2}}|-\rangle+\frac{\cos\epsilon+\sin\epsilon}{\sqrt{2}}|+\rangle,
\end{equation}
where $2\epsilon$ represents a rotation in the azimuthal direction from the state orthogonal to $|i\rangle$ (see appendix in Ref.~\cite{PRX}). This postselection is performed by setting $\delta=0$ in Fig.~\ref{fig:XPM}, and using either a beam-splitter with transmission-reflection ratio given by $(\frac{1+\sin2\epsilon}{2})/(\frac{1-\sin2\epsilon}{2})$ in Fig.~\ref{fig:XPM}(a) or a HWP set to an angle $(\epsilon+45^\circ)/2$ in Fig.~\ref{fig:XPM}(b).

The modular value takes a cumbersome expression, so we evaluate different regimes for simplicity:

\begin{itemize}
\item{$\phi\ll a\ll|\epsilon|\ll1$:} The modular value takes the form
\begin{equation}
R_m\approx\left(1-\frac{a}{2\epsilon}\right)\,e^{i\phi/\epsilon},
\end{equation}
where the cross phase can be amplified by a factor of $1/|\epsilon|\gg1$ with a success probability of $p_{N_{XPM}}\approx\epsilon^2$. Even though the absorption seems to be amplified as well, by choosing a negative value for $\epsilon$ the amplitude of the one-photon Fock state can be slightly amplified with respect to the vacuum eigenstate. 

\item{$\phi\ll|\epsilon|\ll a\ll1$:} If the postselection angle $\epsilon$ can be experimentally smaller than the absorption $a$, the modular value takes a different form,
\begin{equation}
R_m\approx -\left(1-\frac{a}{2\epsilon}\right)e^{-i2\phi/a}.
\end{equation} 
As in the previous case the cross phase is amplified, however it grows proportional to the relative absorption $a$, which at the same time can be converted in a relative significant gain since $a/2|\epsilon|\gg1$. The probability to succeed takes the form $p_{N_{XPM}}\sim\epsilon^2+a^2\alpha^2/4$ in this case.

\item{$a=0$ and $\phi\ll|\epsilon|\ll1$:} If absorption is not present, the cross phase is directly amplified as $R_w\approx e^{i\phi/\epsilon}$ due to postselection. Similar amplification was first introduced in Refs.~\cite{singlephotonnonlinear,CLEO-XPM}. 
\end{itemize}

We have shown here how to induce a WVA-like response plus mitigation of a relative absorption in a two-qubit scenario. Instead of amplifying the parameter of interest ($\phi$ in our case) proportionally to the weak value as in WVA, the modular value completely specifies the final system's state.    

\subsection*{Phase-Amplitude Complementarity}

We explore now a different post-selection on the ancillary qubit: when it is given by a small rotation $\delta$ on the equatorial plane 
from the state orthogonal to $|i\rangle$,
\begin{equation}
|f\rangle=\frac{1}{\sqrt{2}}(|-\rangle+e^{-i\delta}|+\rangle).
\end{equation}
This postselection is performed by setting $\delta\neq0$ in Fig.~\ref{fig:XPM}, and by either replacing the unbalanced BS with a 50/50 BS in Fig.~\ref{fig:XPM}(a), or by setting the HPW to an angle of $22.5^\circ$ in Fig.~\ref{fig:XPM}(b). We evaluate different regimes here as well:
\begin{itemize}
\item{$\phi\ll a\ll|\delta|\ll1$:} The modular value takes the interesting form
\begin{equation}
R_w\approx \left(1-\frac{a}{2}\right)\left(1+\frac{\phi}{\delta}\right)\,e^{i\phi+i 2a/\delta},
\end{equation}
where the effective relative amplitude and the effective phase in the final state $|\alpha_f\rangle$ can be controlled by the corresponding physical conjugate variables, i.e. the total relative absorption and total relative phase are both dependent upon $a$ and $\phi$. Note for example, in a null cross phase scenario ($\phi=0$) an effective controllable cross phase ($2a/\delta$) can still be induced by a small absorption $a$ and post-selection. Also, if absorption is not initially present, $a=0$, such can still be introduced using the cross phase $\phi$ and postselection. We call this behaviour \textit{phase-amplitude complementarity in XPM}, and the process to perform it has a success probability of $p_{N_{XPM}}\approx\delta^2/4$. 

\item{ $\phi\ll|\delta|\ll a\ll1$:} Under the approximation of stronger postselection than experimental absorption, the modular value is given by
\begin{equation}
R_m\approx -\frac{a}{\delta}\,e^{-i2\delta /a}.
\end{equation}  
The complementary behaviour is unidirectional here: the initial absorption $a$ plus postselection determines the outcome independently of the cross phase $\phi$. Besides inducing an effective cross phase much larger than $\phi$, the absorption is \textquotedblleft moved" to the vacuum eigenstate of the coherent state since $a/|\delta|\gg1$, and the probability of success is boosted to $p_{N_{XPM}}\approx(\delta^2/4)[1+\alpha^2(a/\delta)^2]$. Note that the result requires $\delta\neq0$ and $a\neq0$. As opposed to the previous example where $|R_m|\approx1$, this results is obtained because $|R_m|\gg1$.

\end{itemize}


\section{Conclusions}
We have presented the advantages of performing full control on the state of a qubit (system) by local pre- and post-selection of an ancillary qubit to overcome technical limitations or experimental imperfections. A conditional non-linear interaction between both qubits is required, and a modular value for the ancillary qubit is defined which determines the final state of the system qubit. The probability of success of the protocol depends upon the overlapping of the pre- and post-selection states for the ancillary qubit, the magnitude of the modular value, and the initial prepared state of the system qubit. By performing local measurements as pre- and post-selection on an ancillary qubit, experimental imperfections in a non-linear two-qubit gate can be overcome in a probabilistic manner. Even though the protocol relies on the probability of post-selection success, the gate is deterministic in the sense that the quantum state at the output is known for the successful/useful cases.

We applied the protocol to a cross phase modulation scenario, where the qubit-qubit cross effective operation is a non-unitary phase shift gate. By introducing post-selection on the ancillary qubit, a small cross phase can be amplified and relative absorption diminished. In addition, a complementary behaviour where an \textit{effective} cross phase (absorption) is induced by the \textit{directly-induced} absorption (cross phase) is presented. We note that similar advantage due to this phase-amplitude complementary is behind the technique of phase-contrast microscopy, where a phase shift is converted to a brightness shift.


Our postselection-based proposed gate is different that previously introduced CNOT gates based on weak measurements~\cite{BrunGate,Lund}, and we hope it sparks interest for quantum processing of information.     


\appendix
\section{Nonunitary phase shift gate}\label{appendix}
We introduce here an imperfect single-qubit phase shift gate by defining a unwanted absorption. Let's recall first the unitary phase shift gate. This gate introduces a relative phase $\phi$ in a qubit, i.e. $\cos(\theta/2)|0\rangle+e^{i\xi}\sin(\theta/2)|1\rangle\rightarrow \cos(\theta/2)|0\rangle+e^{i(\xi+\phi)}\sin(\theta/2)|1\rangle$. The operation corresponds to a rotation in the Bloch sphere about the z-axis by an angle $\phi$, 
\begin{equation}\label{unitary}
R_z(\phi)=\exp\left(-i\frac{\phi}{2}\,\hat{\sigma}_z\right)=
\begin{bmatrix}
e^{-i\phi/2}&0\\
0&e^{i\phi/2}\\
\end{bmatrix}.
\end{equation}
We consider now a small relative absorption $a$ by replacing $\phi\rightarrow \phi+ia$ in eq.~(\ref{unitary}), where $a\ll1$. The nonunitary rotation can be expressed as
\begin{equation}\label{Kraus1}
R_z^{NU}(\phi)\approx\begin{bmatrix}
e^{-i\phi/2}&0\\ 
0&(1-a)\,e^{i\phi/2}\\
\end{bmatrix},
\end{equation}
and it represents experimental imperfections during the operation of the gate. The renormalized state after such a rotation takes the form
\begin{eqnarray}
|\Psi\rangle&=&\frac{R_z^{NU}|\psi\rangle}{\sqrt{\langle\psi|(R_z^{NU})^\dagger R_z^{NU}|\psi\rangle}}\\&\approx&\frac{\cos(\theta/2)|0\rangle+(1-a)\sin(\theta/2)e^{i(\xi+\phi)}|1\rangle}{1-a\sin^2(\theta/2)},
\end{eqnarray}
and the operation has a success probability given by $p_N=\langle\psi|(R_z^{NU})^\dagger R_z^{NU}|\psi\rangle\approx1-2a\sin^2(\theta/2)$, where $|\psi\rangle$ is the initially prepared state of the system qubit. The operator $\hat{R}_z^{NU}(\phi)$ can alternatively be seen as part of a set of two POVMs, so it is also possible for the qubit to be projected to the state $|1\rangle$ with a small probability $2a\sin^2(\theta/2)$. We call this case a negative result, and we compensate the positive result with the small backaction on the qubit. In this sense, the qubit after the nonunitary rotation of Eq.~(\ref{Kraus1}) is considered to remain in a pure state but the desired rotation $\phi$ is effectively accompanied with a undesired small rotation $\Delta\theta\approx-a\sin\theta$ along the azimuthal direction. The state takes the form,
\begin{equation}
|\Psi\rangle\approx\cos\left(\frac{\theta+\Delta\theta}{2}\right)|0\rangle+e^{i(\xi+\phi)}\sin\left(\frac{\theta+\Delta\theta}{2}\right)|1\rangle.
\end{equation}
 
\begin{acknowledgments} 
J. M.-R. thanks Kristin M. Beck for sharing preliminary experimental results on postselected XPM which motivated this work, Justin Dressel, Andrew Jordan, and Bethany Little for valuable discussions and feedback on the manuscript, and Northrop Grumman Corporation for financial support. 
\end{acknowledgments}



%

\end{document}